\documentclass[fleqn,usenatbib]{mnras}

\usepackage{newtxtext,newtxmath}

\usepackage[T1]{fontenc}
\usepackage{ae,aecompl}

\usepackage{graphicx}	
\usepackage{amsmath}	
\usepackage{amssymb}	
\usepackage{pdflscape} 
\usepackage{multirow}

\title[AGN and non-AGN twins. Spotting the differences]{Spotting the differences between active and non-active twin galaxies on kpc-scales. A pilot study.}

\author[I. del Moral-Castro et al.]{I. del Moral-Castro,$^{1,2}$\thanks{E-mail: imoralc@iac.es}
B. Garc\'ia-Lorenzo,$^{1,2}$
C. Ramos Almeida,$^{1,2}$
T. Ruiz-Lara,$^{1,2}$
\newauthor J. Falc\'on-Barroso,$^{1,2}$ 
S.F. S\'anchez,$^{3}$
P. S\'anchez-Bl\'azquez,$^{4}$
I. M\'arquez$^{5}$
\newauthor and J. Masegosa$^{5}$
\\
$^{1}$Instituto de Astrof\'isica de Canarias, C/ V\'ia L\'actea, s/n, E-38205 La Laguna, Tenerife, Spain\\
$^{2}$Departamento de Astrof\'isica, Universidad de La Laguna, E-38206 La Laguna, Tenerife, Spain\\
$^{3}$Instituto de Astronom\'ia, Universidad Nacional Aut\'onoma de M\'exico, A. P. 70-264, C.P. 04510, M\'exico, D.F., Mexico\\
$^{4}$Departamento de F\'isica Te\'orica, Universidad Aut\'onoma de Madrid, Cantoblanco, E-28049 Madrid, Spain\\
$^{5}$Instituto de Astrof\'isica de Andaluc\'ia (CSIC), Apdo. 3004, 18080, Granada, Spain\\
}

\date{Accepted XXX. Received YYY; in original form ZZZ}

\pubyear{2019}

\begin{document}
\label{firstpage}
\pagerange{\pageref{firstpage}--\pageref{lastpage}}
\maketitle

\begin{abstract}
We present a pilot study aimed to identify large-scale galaxy properties that could play a role in activating a quiescent nucleus. To do so, we compare the properties of two isolated nearby active galaxies and their non-active twins selected from the Calar Alto Legacy Integral Field Area (CALIFA) survey. This pilot sample includes two barred and two unbarred galaxies. We characterise the stellar and ionised gas kinematics and also their stellar content. We obtain simple kinematic models by fitting the full stellar and ionised gas velocity fields and just the approaching/receding sides. We find that the analysed active galaxies present lopsided disks and higher values of the global stellar angular momentum ($\lambda_{R}$) than their non-active twins. This could be indicating that the stellar disks of the AGN gained angular momentum from the inflowing gas that triggered the nuclear activity. The inflow of gas could have been produced by a twisted disk instability in the case of the unbarred AGN, and by the bar in the case of the barred AGN. In addition, we find that the central regions of the studied active galaxies show older stellar populations than their non-active twins. The next step is to statistically explore these galaxy properties in a larger sample of twin galaxies.
\end{abstract}

\begin{keywords}
Galaxy: kinematics and dynamics -- Galaxy: nucleus -- galaxies: active
\end{keywords}



\section{Introduction}

Observational studies indicate that all massive galaxies harbour supermassive black holes at their centres (see \citealt{Ho_2008} for a review) with masses that scale with  parameters of their host galaxy (e.g. \citealt{Ferrarese_2000, Gebhardt_2000, Tremaine_2002, Kormendy_2013}). These evidences for the co-evolution of galaxies and their central black holes could be indicating that all galaxies might experience nuclear activity for at least some part of their evolution, active phases in which the black hole accretes material from the host galaxy and grows \citep{Soltan_1982,Schawinski_2015}. Therefore, unveiling the mechanisms that trigger the active galactic nuclei (AGN) phase is crucial for understanding the formation and evolution of galaxies. 

Traditionally, interactions and mergers have been suggested as the primary responsible for igniting nuclear activity \citep{1977ApJS...33...19A, 1988ApJ...325...74S, 1991ApJ...370L..65B, 1996ApJ...464..641M, Sanchez_2005}, but an important fraction of AGN occur in seemingly undisturbed galaxies (\citealt{Marquez_1999,Marquez_2000, Marquez_2003, Marquez_2004, Sebas_2004, 2009ApJ...691..705G, 2011ApJ...726...57C, 2012ApJ...761...75S, Kocevski_2012, 2013MNRAS.429.2199S}). 

Thus mergers/interactions are not a mandatory process for triggering AGN and alternative mechanisms may be more important than previously thought. In fact, there seems to be a dependence between the prevailing triggering mechanism and the AGN luminosity (e.g \citealt{Hopkins_2008, Treister_2012}). While luminous AGN would be triggered by interaction/mergers (e.g. \citealt{Cris_2011,Bessiere_2012,Chiaberge_2015}), low-luminosity AGN would be driven by secular processes (e.g. \citealt{2011ApJ...726...57C}) such as bar/disk instabilities, spiral structures or weakly nonaxisymetric potentials (see \citealt{Shlosman_1990, Wada_2004, Bournaud_2012, Menci_2014}). 

Common in the nearby universe ($\sim$60\,$\%$ of spiral galaxies having a bar, \citealt{2000ApJ...529...93K, 2009A&A...495..491A}), stellar bars are known to be efficient agents for driving gas to the central kpc (e.g. \citealt{Heller_1994}). However, the link between bars and nuclear activity is ambiguous \citep{1989A&A...217...66A, 1995ApJ...438..604M, 1997ApJ...487..591H, 1997ApJ...482L.135M, Marquez_2000, 2000ApJ...529...93K, 2002ApJ...567...97L, 2009ASPC..419..402H, 2013ApJ...776...50C}. Indeed AGN are also observed in unbarred galaxies,  suggesting that the presence of stable bars are non-essential. 

Many works performed at optical wavelengths through a variety of observational techniques (i.e. narrow-band imaging, long-slit spectroscopy, Fabry-Perot interferometry and more recently integral field spectroscopy) have tried to disentangle the main AGN triggering mechanism in the local universe (e.g. \citealt{Marquez_2003, Marquez_2004, Sebas_2005,Dumas_2007}). Despite these efforts, the main driver for AGN triggering in isolated galaxies still remains unclear. The definition of non-active samples is essential to investigate and identify properties unique to AGN and to ensure that the potential differences are not related to galaxy properties like mass or Hubble type. Due to sample limitations (e.g. number of objects, data homogeneity) to date it has been difficult to make one-to-one comparisons. 

Legacy integral field surveys like CALIFA \citep{2012A&A...538A...8S}, SAMI \citep{SAMI_2012} or MANGA \citep{MANGA_2015}, offer 3D optical information for a large number of galaxies covering a wide range of stellar masses and morphological types. These surveys provide, for the first time, the opportunity of selecting large-scale almost-identical pairs of isolated galaxies differing only in nuclear activity. By comparing the properties of these almost-identical twins we should be able to identify any difference connected to AGN triggering induced by secular processes. 

This is the first of a series of papers exploring possible large scale AGN triggering mechanisms in isolated galaxies. This pilot study is intended to outline the methodology and identify relevant parameters related to AGN triggering to be explored in a larger sample.

This paper is organized as follows. In Section \ref{Sample} we summarize the observations and criteria to select the pilot sample. In Section \ref{sec:kinematic} we explain the methods used to extract the parameters considered in this pilot study. The results and discussion are presented in Section \ref{Results}. The main conclusions are summarized in Section \ref{sec:Conclusions}.

\section{Observations and sample selection}\label{Sample}

\subsection{The CALIFA survey} 

We have used 3D optical data from the third Data Release  \citep{2016A&A...594A..36S} of the Calar Alto Legacy Integral Field Area (CALIFA) survey \citep{2012A&A...538A...8S}. CALIFA comprises more than 800 galaxies in the Local Universe of all Hubble types, observed with the PMAS instrument at the 3.5\,m telescope of the Calar Alto observatory \citep{2005PASP..117..620R} in the PPak mode \citep{2004AN....325..151V, 2006PASP..118..129K}. 
The main CALIFA sample properties are: redshift in the range 0.005 $< z <$ 0.03, angular isophotal diameter 45'' $<$ isoA$_{r}$ $<$ 79.2'', galactic latitude $|b| >$ 20$^\circ$, declination $\delta >$ 7$^\circ$ and -19 > M$_{r}$ > -23. In the third Data Release the fully-reduced data cubes of 667 galaxies became publicly available. 

We used the COMBO data cubes, that are a combination of the V500 (wavelength range 3745-7500\,\AA, $R \sim 850$ at $\sim 5000$\,\AA) and V1200 (wavelength coverage 3700-4840\,\AA, $R \sim 1650$ at $\sim 4500$\,\AA) CALIFA spectral setups. These data cubes include several optical absorption and emission lines, allowing the extraction of both stellar and ionised gas kinematics \citep{2013A&A...558A..43S, 2014MNRAS.444.3961D, Falcon_2015, 2015A&A...573A..59G, Falcon_2017} and also information on the stellar content and stellar populations of the galaxies (e.g. \citealt{2014A&A...570A...6S, Gonzalez_delgado_2015, 2015A&A...583A..60R, 2017A&A...604A...4R}). More details on the observational strategy, data quality, data reduction and statistical properties of the CALIFA sample can be found in \citet{2012A&A...538A...8S,2016A&A...594A..36S} and \citet{2014A&A...569A...1W}.

\subsection{Sample selection}\label{sec:Selection}

From the full CALIFA DR3 sample, \citet{2017A&A...598A..32M} selected 404 galaxies to characterise their 2D morphology. This is the parent sample for this work. Such selection excludes galaxies in pairs/interaction (57), very distorted (9), edge-on (183) and galaxies strongly contamined by field bright stars (5). 

\begin{figure}
\begin{center} 
 \includegraphics[width=0.45\textwidth]{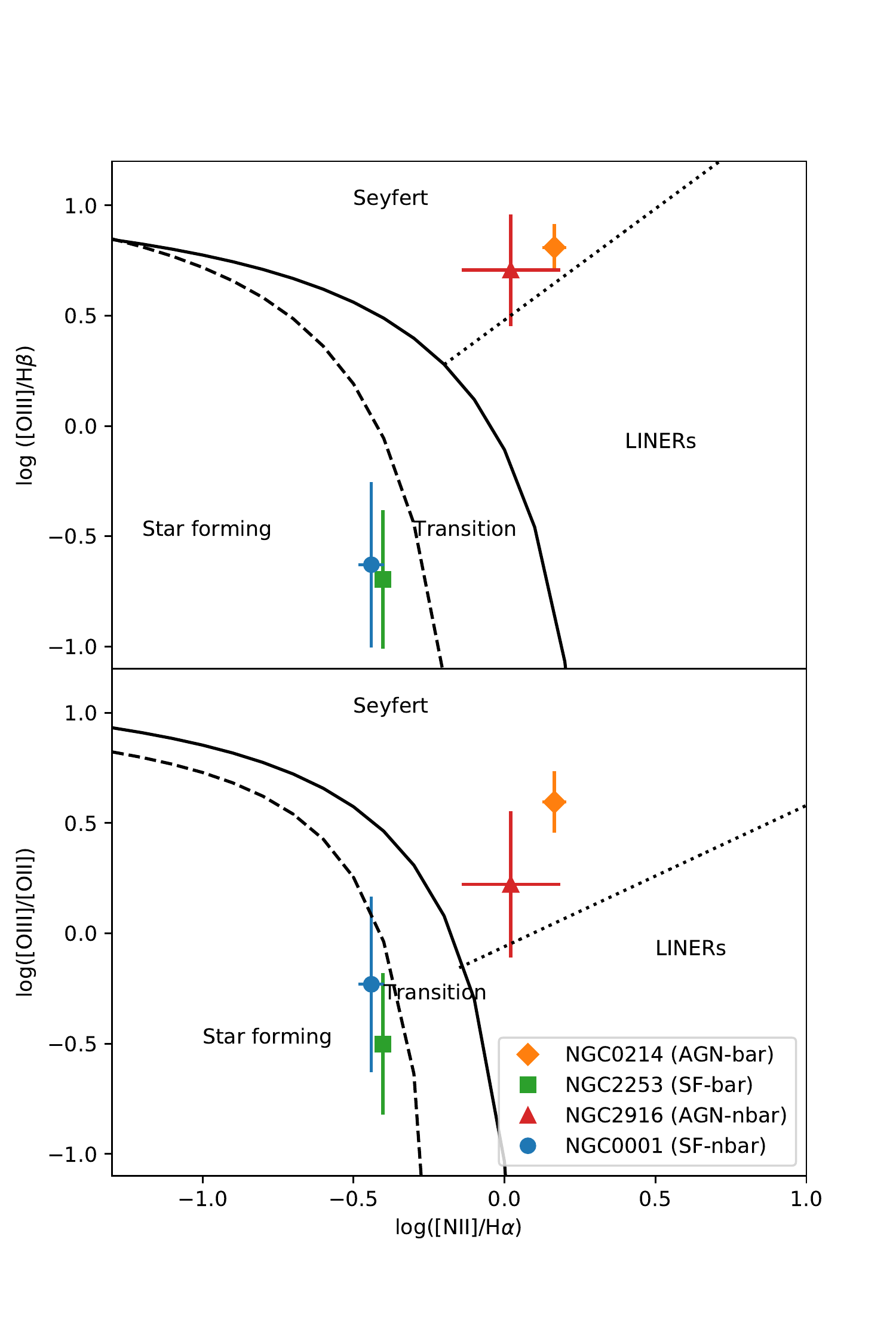} 
 \caption[]{Diagnostic diagrams for the central spectrum of the four galaxies. The lines separating Seyferts, transition objects, LINERs and star-forming galaxies in the top panel are from \citet[][solid curve]{2001ApJ...556..121K}, \citet[][dashed curve]{2003MNRAS.346.1055K} and \citet[][dotted curve]{2006MNRAS.372..961K}. The lines in the bottom panel are from \citet{2010MNRAS.403.1036C}. Error bars correspond to the standard deviation from the mean, calculated from a series of 100 Monte Carlo simulations in order to provide realistic uncertainties.} \label{fig:BPT_diagram_mc}
 \end{center}
\end{figure}

In order to identify AGN we analysed the spectroscopic properties of the ionised gas in the central spaxel of each object following the procedures explained in Section \ref{ssec:velocity_maps}. We consider AGN candidates all galaxies above the line of \citet{2001ApJ...556..121K} and at the left of the Seyfert/LINER division line proposed by \citet{2006MNRAS.372..961K} in the [NII]-BPT diagram \citep{1981PASP...93....5B}. See Fig. \ref{fig:BPT_diagram_mc} for reference of the location of these demarcation lines. Additionally we cross-checked this classification with different databases (e.g. NED and SIMBAD) and previous works (e.g \citealt{2015A&A...573A..59G}). We identify 15 AGN among the 404  galaxies ($\sim$4\% of our parent sample), similar to the AGN fraction reported in nearby galaxy samples (e.g. \citealt{Schawinski_2010, 2014A&A...569A...1W, 2015A&A...573A..59G, Sebas_2018}).

\begin{figure}
    \centering   
    \includegraphics[width=0.45\textwidth]{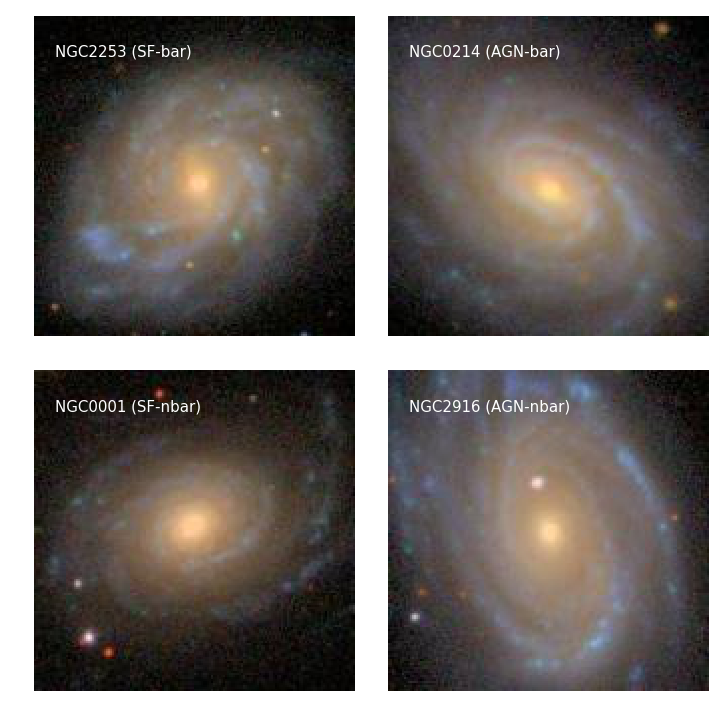}    
    \centering
    \caption[]{Color-composite SDSS images of the four galaxies selected for this pilot study.} 
    \label{fig:galaxies_images}
\end{figure}

From these 15 AGN, we selected the two best matching in redshift, stellar mass, inclination and morphology, but differing in the presence of a bar, resulting NGC0214 (AGN-bar, hereafter) and NGC2916 (AGN-nbar, hereafter). These galaxies are low-luminosity AGN and hence will have low impact AGN feedback on the circumnuclear region, making them ideal targets to look for triggering imprints. We then selected the barred and unbarred galaxies among the remaining 389 objects located at the star-forming region (SF) in the BPT diagram best matching the galaxy properties listed above. The bar length was also taken into account in the selection in order to match the NGC0214 bar length. The selected objects were NGC2253 (SF-bar, hereafter) and NGC0001 (SF-nbar, hereafter). Finally, we confirmed the classification of these four galaxies using the alternative diagnostic diagram [OIII]/[OII] versus [NII]/H$\alpha$ proposed by \citet{2010MNRAS.403.1036C}. 
Figure \ref{fig:BPT_diagram_mc} shows the location of the four selected objects in the two diagnostic diagrams used to assess their nuclear type. Fig. \ref{fig:galaxies_images} shows their broad band SDSS images.

The four selected objects were included in the \citet{2014A&A...568A..70B} sample of non-interacting CALIFA galaxies. Therefore, we adopt their criteria to confirm these galaxies as isolated, that is: no companions having similar systemic velocities (difference smaller than 1000 km\,s$^{-1}$) and SDSS r-band magnitudes (smaller than 2 mag) within a physical radius of 250\,kpc. Table \ref{tab:Sample_information} shows their main characteristics. 

As we already noted, the aim of this pilot study is to outline the methodology and identify relevant parameters related to AGN triggering. However, the four selected galaxies also permit us to explore the large-scale differences due to bars.

\begin{table*}
\begin{center}
\resizebox{18cm}{!} {
\begin{tabular}{ c c c c c c c c c c c c c}
\hline
\hline
Galaxy & CALIFA & Hubble & Distance & Spatial & Stellar  & Magnitude  & Inclination  & Bar & Effective &  SFR & Classification\\
& ID &  type & & scale &  mass & (r band)  & (r band)  & radius & radius & \\
 &  & & (Mpc) & (pc arcsec$^{-1}$) & $\log$ (M/M$_{\odot}$) & (mag)  & (degree) & (arcsec) & (arcsec)  &  (M$_{\odot}$/year) \\
\hline
NGC0214 & 28 & SBbc & 52.8 & 256 & 10.73 & -21.93 & 46 & 17.99 & 18 &   5.37 &  AGN \\
NGC2253 & 147 &  SBbc & 55.6 & 270 & 10.50 & -21.34 & 38 & 14.35 & 15  & 2.19 & SF \\
NGC2916 & 277 & Sbc & 51.0 & 247 & 10.64 & -21.25 & 49 & --- & 26* &  1.51  & AGN \\
NGC0001 & 8 & Sbc & 53.7 & 260 & 10.58 & -21.30 & 51& --- & 12 & 4.47 & SF \\
\hline
\end{tabular}
}
\caption[Galaxy sample]{Basic properties of the four galaxies selected for this pilot study: (1) Galaxy name, (2) CALIFA identifier, (3) Hubble type from CALIFA, (4) distance and (5) spatial scale from the NED database \citep{Theureau_2007}, (6) total stellar mass and (7) Petrosian magnitude from \citet{2014A&A...569A...1W}, (8) inclination based on ellipticity and (9) bar radius from \citet{2017A&A...598A..32M}, (10) effective radius  from \cite{Falcon_2017}, (11) SFR from \citet{2016RMxAA..52..171S} and (12) nuclear type according to BPT diagrams (this work). 

* Note that NGC2916 has a star about 10 arcsec North of the nucleus affecting the effective radius estimation, which ranges from 14 to 26 arcsec in previous works \citep{2014A&A...570A...6S, 2015A&A...573A..59G,Falcon_2017}. The effective radius and SFR  were not  selection criteria for this pilot study.}\label{tab:Sample_information} 

\end{center}
\end{table*}

\section{Data analysis}\label{sec:kinematic} 

In this section we briefly describe the methodology and all the steps carried out for the proper characterisation of the stellar/ionised gas kinematics and stellar content of our sample of galaxies. The byproducts of this analysis not shown in the main body of the text are included in Appendix \ref{sec:kinematic_map}.

\subsection{Kinematics}\label{ssec:velocity_maps}

The stellar kinematics of the galaxies were obtained following the strategy described in \citet{Falcon_2017} and briefly summarized here. We selected all the spaxels with an average signal-to-noise ratio (S/N) larger than 3 in the CALIFA data cubes. We spatially binned the spaxels using the Voronoi 2D binning method for optical IFS data implemented by \citet{2003MNRAS.342..345C} to achieve a minimum S/N of $\sim$20. This value allows us to improve the quality of the spectra in order to reliably estimate  the  first  two  moments  of the line-of-sight stellar velocity  distribution  (LOSVD) while preserving a reasonable spatial resolution. We refer to these Voronoi bins as "voxels" hereafter. About $\sim$80\,$\%$ of the voxels include less than 5 spaxels, with voxels conformed by single spaxels at the central region of the galaxies and larger voxels (> 5 spaxels) at the outer regions.

We used the ``penalised pixel fitting'' (\textsc{pPXF}\footnote{\url{http://www-astro.physics.ox.ac.uk/~ mxc/software/}}) code \citep{2004PASP..116..138C, 2011MNRAS.413..813C} to obtain stellar velocities and stellar velocity dispersions. \textsc{pPXF} fits each voxel spectrum with a combination of stellar templates while masking the emission lines. We used as templates a subset of the \citet{2010MNRAS.404.1639V} models based on the MILES stellar library\footnote{The models are available at \url{http://miles.iac.es}} \citep{2006MNRAS.371..703S, 2011A&A...532A..95F}. Uncertainties estimated through Monte Carlo simulations are $\sim$25\,km s$^{-1}$ on average, for both stellar velocity and stellar velocity dispersion, increasing inside out. 

We measured the emission line properties using the Gas AND Absorption Line Fitting\footnote{\url{http://star-www.herts.ac.uk/\~sarzi/PaperV_nutshell/PaperV_nutshell.html}}
(\textsc{GANDALF}) code  \citep{2006MNRAS.366.1151S, 2006MNRAS.369..529F}. GANDALF models the emission lines present in the spectra as additional Gaussian profiles to the best stellar template determined by pPXF. Only one Gaussian per emission line was considered to obtain the ionised gas velocity and velocity dispersion. We assumed that the stellar kinematics in each voxel are rather smooth and do not change significantly from spaxel to spaxel within each voxel \citep{2015A&A...573A..59G}. We then  fixed the values of stellar velocity and $\sigma$ to the best values of the voxel for the calculation of each individual spaxel therefore reducing the number of degrees of freedom. In this way, no spatial binning was considered for the ionised gas kinematics. Errors were estimated via 100 Monte Carlo simulations. To study the kinematics of the ionised gas we focused on the H$\alpha$ emission due to its higher S/N and more extended emission as compared with other emission lines. The values obtained for both stellar and ionised gas components are in agreement with previous measurements of CALIFA galaxies \citep{2016RMxAA..52...21S,Falcon_2017}.

Appendix \ref{sec:kinematic_map} presents the velocity and velocity dispersion maps of the stellar and ionised gas components for the four selected galaxies. In all cases, the velocity fields resemble rotating disks, with velocity dispersions decreasing outwards (see Fig. \ref{fig:profile_gaseous_dispersion}). High velocity dispersion values in the outer parts are attributable to the poor S/N of the spaxels.  

\subsection{Kinematic model}\label{Modeling}

We use the \textsc{DiskFit}\footnote{\url{https://www.physics.queensu.ca/Astro/people/Kristine_Spekkens/diskfit/}} software package \citep{2007ApJ...664..204S, 2015arXiv150907120S} to model the velocity fields obtained in the previous section for both the stars and ionised gas.
\textsc{DiskFit} has been described and used extensively in previous works 
(e.g. \citealt{2016MNRAS.456.1299C, 2017MNRAS.469.3541P}). In particular, it has been successfully applied to CALIFA data in \citet{2015MNRAS.451.4397H}. \textsc{DiskFit} is a non-parametric algorithm which fits a physically motivated axisymmetric or non-axisymmetric model to the velocity field. The best-fitting parameters are derived by minimising $\chi^{2}$. For this work we assume the simplest model applicable to the four galaxies, a flat axisymmetric disk. The velocity field is given by: 
\begin{eqnarray}
    V_{model} (R) &=& V_{sys} +  V_{t} (R) \sin i\cos (\theta) 
\end{eqnarray}
where $Vsys$ is the systemic velocity, V$_{t} (R) $ is the rotation velocity, $\theta$ is the angle relative to the kinematic major axis and i is the disc inclination. \textsc{DiskFit} determines the position angle of the disc (PA), the kinematic centre, the ellipticity ($\epsilon$), the inclination  and the rotation curve. The parameters can be fixed at an input value or left free.

This simple model was fitted to the full velocity field, but also to the approaching and receding sides of the galaxies separately to assess possible large-scale kinematic distortions. In a purely  rotating disk, the total and sided kinematic parameters should be coincident. Any discrepancies would be suggesting a kinematic lopsidedness. As initial estimations for the model we used the inclination and PA derived from photometric decomposition \citep{2017A&A...598A..32M}. The initial value for the kinematic centre was taken at the peak of a continuum image recovered by integrating the spectra in the mostly emission-line free wavelength range 5700-6050\,\AA\, when fitting the entire velocity field. For the fit of the receding and approaching sides we fixed the kinematic centre to the position resulting from the model of the whole velocity field.

To fit a rotating disk, DiskFit considers different elliptical rings on the observed velocity field up to a selected radius. As the typical spatial resolution of the CALIFA data is 2.5 arcsec \citep{2016A&A...594A..36S}, we sampled the velocity fields with rings separated by 3 arcsec up to a distance of 30 arcsec ($\sim$8 kpc at the average distance of the galaxies). We followed this procedure for the stellar and ionised gas velocity fields of the four analized galaxies. Only in the case of the stellar velocity field of NGC0001 the fit was performed up to 15 arcsec due to the poor spatial resolution in the outer parts (large voxels). Uncertainties on the model output have been estimated through 1000 bootstrap realizations \citep{2010MNRAS.404.1733S} of each fit. Errors are always higher for the stellar kinematics due to the loss of spatial resolution caused by the binning scheme adopted to increase the S/N of the stellar component (see Section \ref{ssec:velocity_maps}).

In order to discard deviations introduced by any possible AGN blurring in the central region of the galaxies, mainly affecting the ionised gas, we also fit the simple disk model to the ionised gas velocity fields masking the central region of the galaxies. This way we can also focus on the large-scale disk kinematics, avoiding any deviations due to e.g. the presence of the bar in the barred galaxies. We considered masks of 6 arcsec in diameter (about twice the CALIFA PSF), 12 arcsec (about four times the CALIFA PSF and $\sim$2/3 of the effective radius, R$_{e}$), and 18 arcsec (larger than R$_{e}$ and bar sizes). The resulting kinematic parameters are almost the same when we use the two larger masks (outer masks). When we compare the results from the 6 arcsec mask (inner mask) and without mask we do not find significant differences. The differences are found when comparing the outer and inner masks results (see Tables \ref{tab:big_table} and \ref{tab:big_table_mask} and Section \ref{inclina}).
 
Appendix \ref{sec:kinematic_map} presents the kinematic models obtained with \textsc{DiskFit} when considering the whole, receding and approaching sides of the velocity field, the residuals after subtracting the models to the velocity fields and the derived rotation curves. Table \ref{tab:big_table} summarizes the photometric (from \citealt{2017A&A...598A..32M}) and kinematic parameters derived. 

Overall the stellar and ionised gas velocity fields are well reproduced by a pure rotating disc at these scales. The mean of the absolute values of the residuals are of the order of $\sim$20\,km s$^{-1}$ and $\sim$ 10\,km s$^{-1}$ for the stellar and ionised gas components, respectively. In general, the global kinematic parameters for both components are in good agreement within the uncertainties (see Table \ref{tab:big_table}), with differences always smaller than 3$^{\circ}$ and 4$^{\circ}$ for PA and inclination, respectively. We comment further on this in Section \ref{inclina}.

\subsection{Angular momentum}

We have calculated the luminosity-weighted stellar and gas angular momentum per unit mas (see \citealt{Emsellem_2007,Emsellem_2011}) through:

\begin{eqnarray}
    \lambda_{R} \equiv \frac{\langle R | V | \rangle}{\langle R \sqrt{V^{2} + \sigma^{2}} \rangle} = \frac{\displaystyle\sum_{i=1}^{N_{p}} F_{i} R_{i} |V_{i}| }{\displaystyle\sum_{i=1}^{N_{p}} F_{i} R_{i} \sqrt{V_{i}^{2} + \sigma_{i}^{2}} }
\end{eqnarray}
where R$_{i}$ is the galactocentric radius, F$_{i}$ is the flux, V$_{i}$ is the velocity, $\sigma_{i}$ is the velocity dispersion of the $i$th spatial bin and $N_{p}$ is the number of voxels (for the stellar component) or spaxels (for the gaseous component).  The value of $\lambda_{R}$ is close to unity when perfectly ordered rotation dominates over the local velocity dispersion. 

We have derived the radial variation of $\lambda_{R}$ as well as the integrated value within 1R$_{e}$, for both the stars and ionised gas. We have generated 100 Monte Carlo simulations in order to compute uncertainties. Fig \ref{fig:angular_momentum} shows the stellar and ionised gas angular momentum computed within 1R$_{e}$ versus the ellipticity (see also Table \ref{tab:angular_momentum_table}). The values obtained for the stellar angular momentum are in agreement with previous works for similar galaxies (see figure 1 in \citealt{Falcon_2015} or figure 15 in \citealt{2016Cappellari}). The four galaxies are classified as fast rotators according to the separation introduced by  \citet{Emsellem_2011}. Fig. \ref{fig:angular_momentum_profiles} shows the radial profiles of the stellar and ionised gas $\lambda_{R}$ for the four galaxies calculated from the global, approaching and receding sides of the velocity field.

\begin{landscape}
\begin{table}
\begin{center}
\begin{tabular}{c c c c c c c c c c c c}
\hline
\hline
& & Morphology & & & Gas kinematics &  & & & Stellar kinematics & \\
\hline
& & Total  &  & Approaching & Receding & Total & & Approaching & Receding & Total\\
\hline
\multirow{ 3}{*}{NGC0214}  & PA (deg) & 57.62 $\pm$ 0.29 & & 55.40 $\pm$ 0.51 & 55.45 $\pm$ 0.44 & 55.48 $\pm$ 0.19 & & 55.34 $\pm$ 2.24 & 55.36 $\pm$ 1.76 & 55.34 $\pm$ 1.24\\

\multirow{ 3}{*}{\small{(AGN - bar)}} & incl (deg) & 46.02 $\pm$ 1.75 & & 46.07 $\pm$ 1.77 & 49.31 $\pm$ 1.29 & 44.77 $\pm$ 0.90 & & 59.71 $\pm$ 7.99 & 40.94 $\pm$ 6.37 & 45.61 $\pm$ 7.46\\

& Vsys (km s$^{-1}$) & - & & 4487.15 $\pm$ 2.10 & 4477.19 $\pm$ 1.35 & 4479.79 $\pm$ 1.81 & & 4497.31 $\pm$ 4.14 & 4491.64 $\pm$ 2.81 & 4486.84 $\pm$ 3.16\\
\hline
\multirow{ 3}{*}{NGC2253} & PA (deg) & 129.42 $\pm$ 0.29 & &  117.07 $\pm$ 0.69 & 117.70 $\pm$ 0.70 & 117.43 $\pm$ 0.41 & & 120.11  $\pm$ 1.89 & 118.84 $\pm$ 0.96 & 118.69 $\pm$ 0.82\\

\multirow{ 3}{*}{\small{(SF - bar)}} & incl (deg) & 38.00 $\pm$ 1.19 & & 34.64 $\pm$ 4.22 & 38.07 $\pm$ 3.35 & 33.76 $\pm$ 3.29 & & 46.13 $\pm$ 7.21 & 38.49 $\pm$ 5.85 & 35.58 $\pm$  6.31\\

& Vsys (km s$^{-1}$) & - & & 3534.84 $\pm$ 1.95 & 3529.59 $\pm$ 2.19 & 3529.69 $\pm$ 1.66 & & 3527.95  $\pm$ 2.25 & 3521.71 $\pm$ 1.84 & 3523.83 $\pm$ 1.55\\
\hline
\multirow{ 3}{*}{NGC2916} & PA (deg) & 15.45 $\pm$ 0.17 & &  16.39 $\pm$ 0.52 & 16.45 $\pm$ 0.84  & 16.80 $\pm$ 0.19 & & 14.62  $\pm$ 1.19 & 15.07 $\pm$ 1.00  & 15.12 $\pm$ 0.69\\

\multirow{ 3}{*}{\small{(AGN - nbar)}} & incl (deg) & 49.32 $\pm$ 1.97 & & 56.08 $\pm$ 1.54 & 50.84 $\pm$ 2.47 & 50.91 $\pm$ 1.43  & & 49.06 $\pm$ 7.19 & 49.91 $\pm$ 3.71 & 47.88 $\pm$  2.58\\

& Vsys (km s$^{-1}$) & - & & 3672.12 $\pm$ 2.62 & 3664.34 $\pm$ 2.46  & 3667.82 $\pm$ 4.16 & & 3691.22  $\pm$ 2.62 & 3679.28 $\pm$ 1.48  & 3680.57 $\pm$ 1.65\\
\hline
\multirow{ 3}{*}{NGC0001} & PA (deg) &  96.52 $\mp$ 1.13 & & 115.84 $\pm$ 0.72 & 116.07 $\pm$ 1.23 & 115.91 $\pm$ 0.53 & & 113.07 $\pm$ 2.52 & 113.25 $\pm$ 1.45 & 113.35 $\pm$ 1.10\\

\multirow{ 3}{*}{\small{(SF - nbar)}} & incl (deg) & 51.35 $\pm$ 2.13 & & 42.91 $\pm$ 1.61 & 43.13 $\pm$ 3.00 & 40.98 $\pm$ 1.76  & & 30.39 $\pm$ 11.59 & 40.18 $\pm$ 6.33 & 37.23 $\pm$  6.52 \\

& Vsys (km s$^{-1}$) & & & 4486.20 $\pm$ 1.89 & 4481.03 $\pm$ 2.09 & 4483.75 $\pm$ 0.67  & & 4479.83 $\pm$ 2.98 & 4476.97 $\pm$ 1.87 & 4481.53 $\pm$  1.48\\
\hline
\end{tabular}
\caption[DiskFit gas]{Photometric and kinematic parameters of the analysed galaxies. Morphological parameters are from \citet{2017A&A...598A..32M}. Ionised gas and stellar kinematic parameters were derived from fitting a simple rotating disk to the whole (labelled as Total), approaching and receding sides of the velocity field.}\label{tab:big_table} 
\end{center}
\end{table}

\begin{table}
\begin{center}
\begin{tabular}{c c c c c c c c}
\hline
\hline
& & & Gas kinematics (6 arcsec mask) & & &  Gas kinematics (12 arcsec mask)\\
\hline
& &  Approaching & Receding & Total & Approaching & Receding & Total \\
\hline
\multirow{ 3}{*}{NGC0214}  & PA (deg) & 55.36 $\pm$ 0.41 & 55.55 $\pm$ 0.26 & 55.50 $\pm$ 0.16 & 55.38 $\pm$ 0.34 & 55.58 $\pm$ 0.29 & 55.51 $\pm$ 0.20\\

\multirow{ 3}{*}{\small{(AGN - bar)}} & incl (deg) & 46.03 $\pm$ 1.50 & 49.53 $\pm$ 1.31 & 44.73 $\pm$ 0.66 & 47.40 $\pm$ 1.29 & 48.07 $\pm$ 2.15 & 44.13 $\pm$ 0.74\\

& Vsys (km s$^{-1}$) & 4486.77 $\pm$ 1.57 & 4476.60 $\pm$ 2.06 & 4479.51 $\pm$ 1.44 & 4483.46 $\pm$ 1.40 & 4473.15 $\pm$ 2.59 & 4477.73 $\pm$ 0.54\\
\hline

\multirow{ 3}{*}{NGC2253} & PA (deg) & 116.76 $\pm$ 0.78 &  117.65 $\pm$ 0.71 & 117.41 $\pm$ 0.38   & 116.78 $\pm$ 0.65 & 117.59  $\pm$ 0.39 & 117.50 $\pm$ 0.48 \\

\multirow{ 3}{*}{\small{(SF - bar)}} & incl (deg) & 33.81 $\pm$ 3.91 & 37.03 $\pm$ 3.52 & 33.52 $\pm$ 3.59  & 35.91 $\pm$ 3.88 & 33.49 $\pm$ 3.54 & 33.89 $\pm$ 2.58 \\

& Vsys (km s$^{-1}$) & 3534.82 $\pm$ 2.42 & 3529.01 $\pm$ 2.40 & 3529.46 $\pm$ 1.34    & 3533.03  $\pm$ 2.64 & 3527.97 $\pm$ 2.10 & 3528.56 $\pm$ 0.49\\
\hline

\multirow{ 3}{*}{NGC2916} & PA (deg) & 16.65 $\pm$ 0.42 &  16.25 $\pm$ 0.47 & 16.61 $\pm$ 0.53   & 16.57 $\pm$ 0.41 & 16.26  $\pm$ 0.53 & 17.05 $\pm$ 0.26 \\

\multirow{ 3}{*}{\small{(SF - bar)}} & incl (deg) & 55.91 $\pm$ 1.62 & 50.50 $\pm$ 2.96 & 48.77 $\pm$ 2.10 & 56.26 $\pm$ 1.72 & 49.53 $\pm$ 2.90 & 49.89 $\pm$ 2.86\\

& Vsys (km s$^{-1}$) & 3672.50 $\pm$ 2.56 & 3665.25 $\pm$ 3.05 & 3667.35 $\pm$ 2.68    & 3670.97  $\pm$ 2.83 & 3663.79 $\pm$ 3.29 & 3668.14 $\pm$ 2.37\\
\hline

\multirow{ 3}{*}{NGC0001} & PA (deg) & 115.82 $\pm$ 0.34 &  115.96 $\pm$ 0.42 & 115.85 $\pm$ 0.17   & 115.56 $\pm$ 0.29 & 116.08  $\pm$ 0.27 & 115.82 $\pm$ 0.15 \\

\multirow{ 3}{*}{\small{(SF - bar)}} & incl (deg) & 42.23 $\pm$ 1.22 & 42.08 $\pm$ 2.09 & 40.95 $\pm$ 1.87 & 42.53 $\pm$ 1.33 & 41.30 $\pm$ 1.65 & 40.51 $\pm$ 1.47\\

& Vsys (km s$^{-1}$) & 4484.98 $\pm$ 1.67 & 4482.13 $\pm$ 1.70 & 4483.69 $\pm$ 0.54    & 4485.28  $\pm$ 1.86 & 4482.28 $\pm$ 1.67 & 4483.47 $\pm$ 0.49\\
\hline
\end{tabular}
\caption[DiskFit gas]{Kinematic parameters derived from fitting a simple rotating disk to the ionised gas velocity fields when masking the central region with a 6 and 12 arcsec diameter aperture (see text).}\label{tab:big_table_mask} 
\end{center}
\end{table}
\end{landscape}

\subsection{Stellar populations}\label{Stellar population method}

We characterised the stellar populations of the galaxies following the methodology used and tested in  \citet{2011MNRAS.415..709S, 2014A&A...570A...6S} and  \citet{2015A&A...583A..60R, 2017A&A...604A...4R}. Here we highlight the main steps:

\begin{enumerate}
\item We spatially binned the spectra in a similar fashion to what we did to study the stellar kinematics but with a S/N goal of 30, as reliable stellar populations analysis requires larger S/N values. We checked that the results are not contingent upon the choice of the minimum S/N by repeating the analysis with different constraints (30, 45 and 60). We prefer to stick to S/N=30 to preserve the best possible spatial resolution for this analysis.

\item We obtained pure absorption spectra after subtracting the emission lines fitted with \textsc{pPXF} and \textsc{GANDALF} (see Section \ref{ssec:velocity_maps}). The wavelength range was limited to 3800\,\AA{} - 5900\,\AA\ where most of the spectral features sensitive to the stellar  populations are located.

\item We use \textsc{STECKMAP}\footnote{\textsc{STECKMAP} is a public tool and can be downloaded at \url{http://astro.u-strasbg.fr/~ocvirk/}} (STEllar Content and Kinematics via Maximum A Posteriori likelihood; \citealt{2006MNRAS.365...46O, 2006MNRAS.365...74O}) to study the stellar content from the pure absorption spectra. Here we use the new set of MILES models based on the BaSTI \citep{Pietrinferni_2004} isochrones \citep{Vazdekis_2016}.
\end{enumerate}

The typical \textsc{STECKMAP} outputs are the stellar age distribution, the age-metallicity relation and the LOSVD. However, we fix the LOSVD to the values derived from \textsc{pPXF} to avoid stellar dispersion-metallicity degeneration (see Appendix B of \citealt{2011MNRAS.415..709S}). In addition, we can obtain values of age and metallicity per spectrum weighting by light (LW) or mass (MW). Errors for these quantities are computed by means of 25 Monte Carlo simulations, which have been proven to provide reasonable errors in previous works \citep{Seidel_2015,2015A&A...583A..60R,2017A&A...604A...4R}.

Figures \ref{fig:age_Tomas_arcs} and \ref{fig:metallicity_Tomas_arcs} show the average radial variation of age and metallicity. In general, the mass-weighted age gradients are flatter than the luminosity-weighted, suggesting the presence of an underlying old stellar population ($\sim$10\,Gyr). This behaviour is in agreement with previous studies based on CALIFA galaxies \citep{2014A&A...570A...6S, 2016MNRAS.456L..35R, 2017MNRAS.470L.122P}, which showed a high percentage of old stellar populations at all radii. We find that metallicity decreases with radius (see Fig. \ref{fig:metallicity_Tomas_arcs}), also previously reported for stellar populations studies of CALIFA galaxies (e.g. \citealt{
2014A&A...570A...6S, Gonzalez_delgado_2015}).We further comment on Figures \ref{fig:age_Tomas_arcs} and \ref{fig:metallicity_Tomas_arcs} in Section \ref{sec:effects_stellar_population}, focusing on the LW profiles. In Appendix \ref{sec:kinematic_map} we show the mean log(age)/metallicity maps of the four galaxies.

Using the stellar age distribution from \textsc{STECKMAP} to unveil how different star formation epochs influenced each galaxy, we have also analysed the spatial and radial distribution of three different stellar sub-populations based on their age. We adopted the same thresholds of \citet{Seidel_2015} for a sample of nearby galaxies:  young (age $\lesssim$1.5\,Gyr), intermediate (1.5\,Gyr $\lesssim$ age $\lesssim$ 10\,Gyr) and old ( $\gtrsim$ 10\,Gyr). 
Fig. \ref{fig:maps_profiles_SFH_Tomas} shows the maps of the LW contribution of each stellar population in each spatial element of the galaxies and the radial profiles. Overall, we find a lower contribution of young stellar populations in the central region of the AGN than in the non-active galaxies (see Section \ref{sec:effects_stellar_population} for more details).

\section{Results and discussion}\label{Results}

In order to identify key large-scale parameters that could be connected with AGN triggering, in the following sections we mainly focus on the comparison between the AGN and the SF galaxies. The exception is Section \ref{dispersion}, in which we confront the results for the barred vs unbarred galaxies first. We refer to Appendix \ref{sec:kinematic_map} for comments on individual objects.

\subsection{Velocity dispersion radial profiles}\label{dispersion}

\subsubsection{Barred vs unbarred twins}

\begin{figure}
    \centering   
   \includegraphics[width=0.45\textwidth]{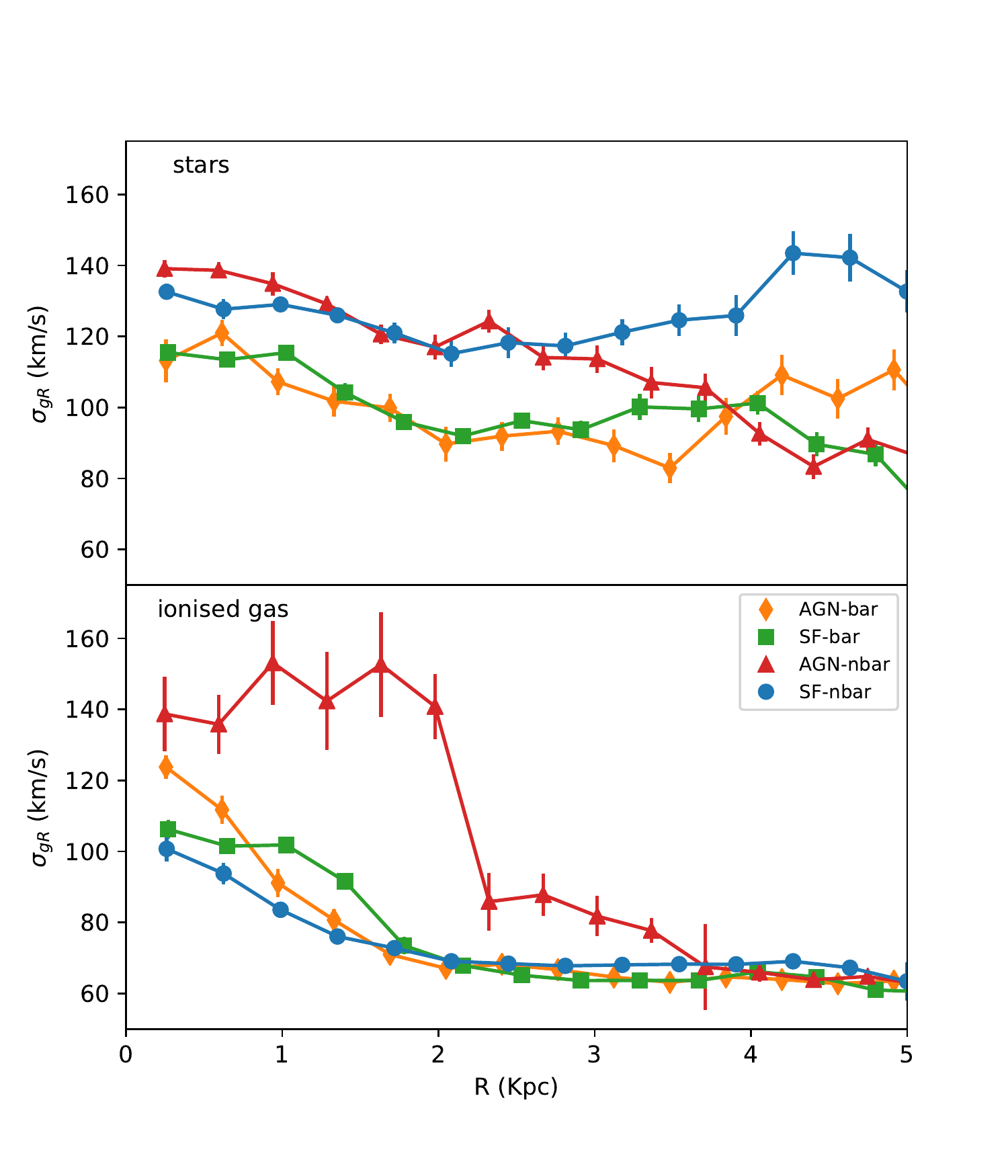}
    \caption{Radial profiles of the stellar and ionised gas velocity dispersions for the inner 5\,kpc of the galaxies. Error bars correspond to rms/$\sqrt[]{N}$ , being N the number of points averaged at each radius.}
    \label{fig:profile_gaseous_dispersion}
\end{figure}

Figure \ref{fig:profile_gaseous_dispersion} shows the variation of the velocity dispersion as a function of galactocentric distance for the stars ($\sigma_{sR}$) and the ionised gas ($\sigma_{gR}$). In the central 3\,kpc of the galaxies, the barred and unbarred twins show similar stellar velocity dispersion radial profiles, with lower $\sigma_{sR}$ for the barred twins. At larger distances, $\sigma_{sR}$ for the unbarred AGN decreases to reach the range of values of the barred twins, while its non-active twin increases $\sigma_{sR}$ outwards. However, the latter is likely associated to the low S/N. 

$\sigma_{sR}$ arises from the combination of different structural components (i.e. bulge, bar, disk) integrated along the line-of-sight. In general bulges are dynamically hotter than bars, and disks are dynamically colds. This means that bulges are  pressure supported systems, with relatively larger velocity dispersions than velocity amplitudes, i.e. V/$\sigma_{s}$ < 1, while disks are rotationally supported systems with V/$\sigma_{s}$ > 1. The relative contribution of these components explains the observed difference in the measured $\sigma_{sR}$ when comparing the barred and unbarred twins. The unbarred galaxies have more massive bulges than their barred twins ($M_{bulge}$/10$^{10}$M$_{\odot}$ = 0.92 $\pm$ 0.09 for the AGN-nbar, 0.58 $\pm$ 0.01 for the AGN-bar, 1.68 $\pm$ 0.2 for the SF-nbar and 0.46 $\pm$ 0.01 for the SF-bar; \citealt{2017ApJ...848...87C}), and therefore a larger contribution from the dynamically hottest component to the velocity dispersion. In the framework of the central black-hole and host galaxy co-evolution (e.g. \citealt{Kormendy_2013}) less massive bulges and lower stellar velocity dispersion would translate into barred galaxies harbouring less massive black holes than their unbarred twins. This suggests that the selected barred galaxies would have followed slightly different evolutionary paths than their unbarred twins, in spite of having a similar kpc-scale morphological appearance.

\subsubsection{AGN vs SF twins}

Regarding the ionised gas component, the two active galaxies show larger central velocity dispersions than their non-active twins in the central kpc, suggesting that the AGN would be disturbing the circumnuclear gas. $\sigma_{gR}$ reaches similar values for the four galaxies at larger galactocentric distances. Note that the larger $\sigma_{gR}$ measured for the AGN-nbar is more attributable to the low H$\alpha$ emission between 1 and 3\,kpc than to a real behaviour.

\subsection{Morpho-kinematic parameters}\label{inclina}

The differences between the kinematic axes derived from the approaching, receding or global velocity distribution and the morphological orientations (see Table \ref{tab:big_table}) for the four analysed galaxies are smaller than 20$\degr$ for both the stellar and ionised gas components. Our findings agree with previous results for similar nearby galaxies, where photometric and kinematic axes are found to be well-aligned in a large fraction (>80\,\%) of the objects (e.g. \citealt{2014A&A...568A..70B} and references therein). Comparing angles in Table \ref{tab:big_table}, we note that the morpho-kinematic alignment in the active galaxies is better than 2.5$\degr$, while their non-active twins present misalignments larger than 10$\degr$ ($\sim$12$\degr$ for NGC\,2253 and $\sim$19.5$\degr$ for NGC\,0001). However, the presence of non-axisymmetric structures (e.g. bars, spiral arms, etc.) makes it difficult to estimate the photometric orientation \citep{2006MNRAS.369..529F, 2011MNRAS.414.2923K, 2014A&A...568A..70B, Mitchell_2018}. Indeed, the most recent photometric positions angles reported for CALIFA galaxies \citep{2018MNRAS.477..845G} reduce the morpho-kinematic misalignments to less than 8$\degr$ for the four objects.

For both the stellar and ionised gas components, the orientation of the kinematic axes derived from the approaching and receding sides of the velocity field are in good agreement, with differences smaller than 1.5$\degr$ in the four galaxies. Therefore, we do not find any evidence of internal kinematic misalignments in any of the galaxies. 

The uncertainties in the determination of the disk inclination from the global, approaching and receding sides of stellar velocity field prevent to derive any conclusion from the comparison between twins, since the differences are smaller than the errors. In the case of the ionised gas component, the differences in disk inclination obtained from the approaching and receding sides are smaller than the errors for the two SF galaxies (i.e. -3.4$\pm5.4\degr$ for NGC2253 and -0.2$\pm3.4\degr$ for NGC0001), but slightly larger than the errors for the two AGN (i.e. 3.2$\pm2.2\degr$ for NGC0214 and 5.2$\pm2.9\degr$ for NGC2916), suggesting and internal twist in the ionised gas disk of the active galaxies.
In the case of NGC0214 the difference becomes smaller than the error (-0.7$\pm2.5\degr$) when the inclination is estimated using ionised gas velocities at large galactocentric distances (i.e. masking the inner 12 arcsec), where the disk dominates the light distribution and kinematics. This behaviour suggests that the simple rotating disk model might need an additional contribution, most probably the bar, to properly fit the observed velocity field in the inner 12 arcsec of NGC0214. We do not do it here because a more complex modelling is beyond the scope of this paper. In the case of NGC2916 the difference between the inclinations estimated from the approaching and receding sides of the velocity field remains when we mask the inner 12 arcsec (6.7$\pm3.4\degr$) suggesting a twisted disk. 
Therefore, we report an internal twist in the ionised gas disk of the unbarred AGN that is not evident in either of its twins. This internal twist in NGC2916 could be the imprint of the disk instability needed to promote the inflow of gas from large scales to the inner region and trigger the AGN. This twist would not be necessary in the barred AGN because the bar itself can efficiently drive gas inwards. The proposed scenario needs to be confirmed using a statistically significant sample of unbarred twins differing in nuclear type. 

\subsection{Stellar and ionised gas angular momentum}\label{angular_momentum}

\begin{figure*}
    \centering   
    \includegraphics[width=0.7\textwidth]{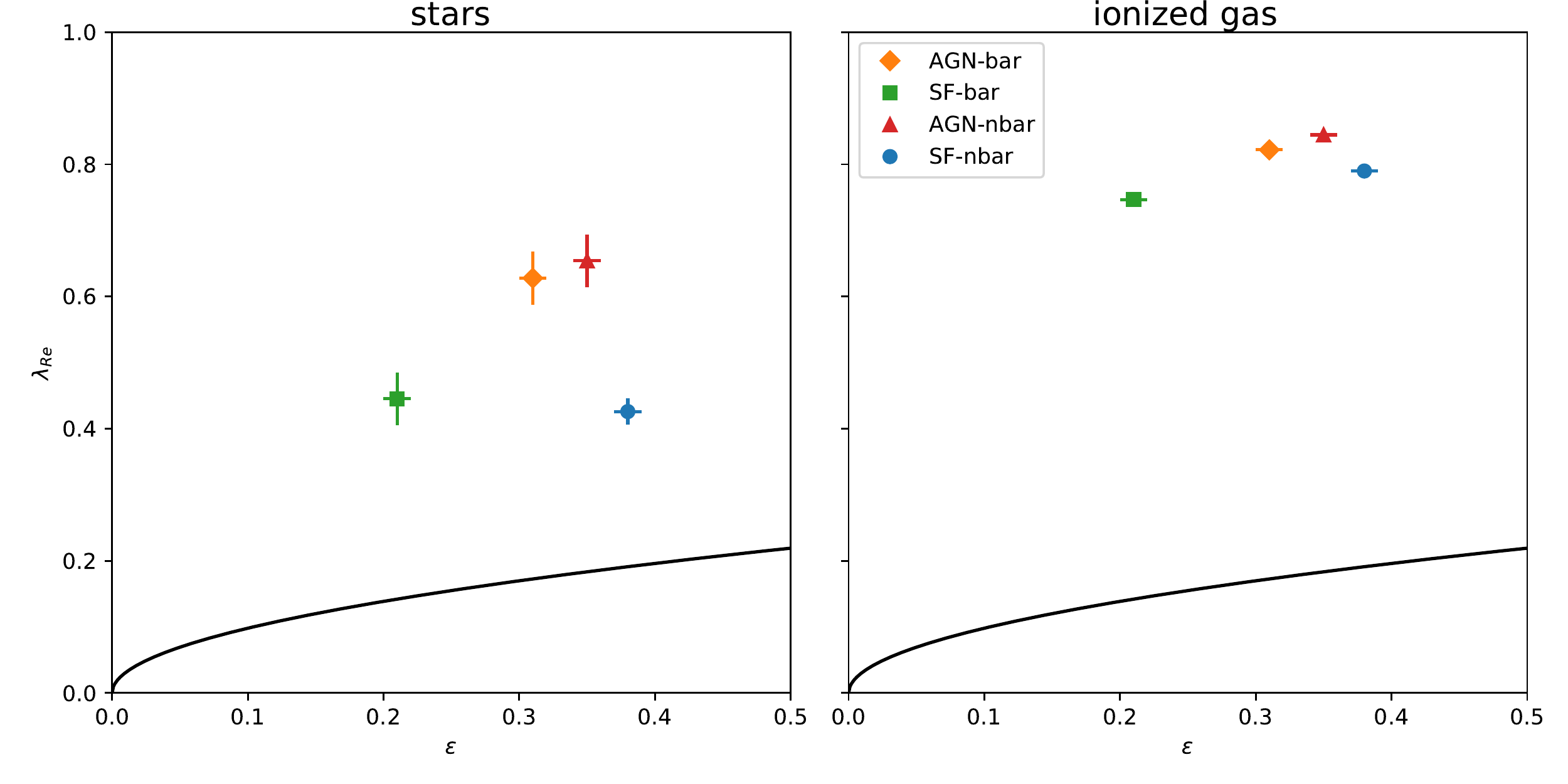}    \caption{Angular momentum ($\lambda_{R_{e}}$) of the  stars (left) and gas (right) versus the ellipticity ($\epsilon$, Table \ref{tab:big_table}) for the four galaxies considered here. The solid line indicates the separation between fast and slow rotators from \citet{Emsellem_2011}.}
    \label{fig:angular_momentum}
\end{figure*}

We find practically the same values of the angular momentum (see Table \ref{tab:angular_momentum_table} and Fig. \ref{fig:angular_momentum}) when we compare the barred and unbarred galaxies, in agreement with previous works \citep{Emsellem_2011, Falcon_2015}. However we find that the two AGN show larger stellar and ionised gas $\lambda_{Re}$ than their non-active twins. Moreover, the differences between the stellar and ionised gas $\lambda_{Re}$ are smaller for the AGN than for their non-active twins ($\Delta \lambda_{Re} \sim 0.19$ for AGN and $\Delta \lambda_{Re} \sim 0.33$ for SF). 

\begin{figure*}
    \centering   
    \includegraphics[width=.8\textwidth]{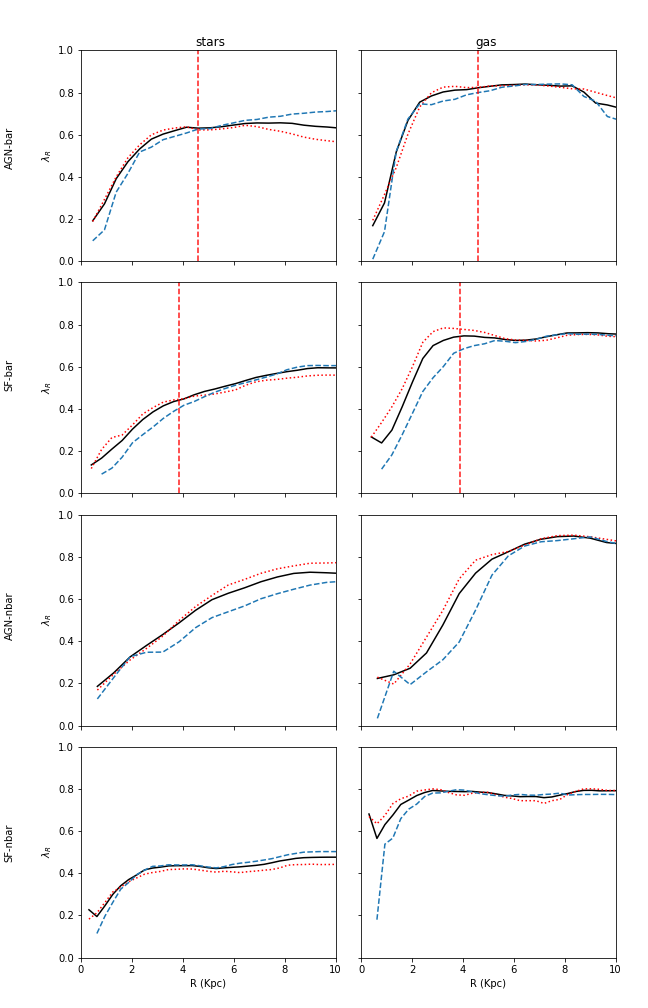}  
    \caption{Radial profiles of the stellar (left) and ionised gas (right) angular momentum ($\lambda_{R}$) for the four galaxies. The solid black, dashed blue and dotted red lines correspond to the total, approaching and receding radial profiles. The dashed red vertical lines indicate the radius of the bars.}
    \label{fig:angular_momentum_profiles}
\end{figure*}

Figure \ref{fig:angular_momentum_profiles} shows radial profiles with larger values of $\lambda_{R}$ for the AGN than their SF twins beyond the central 2\,kpc, at least for the stellar component (left panels in Fig. \ref{fig:angular_momentum_profiles}). Interestingly, we find larger differences between the AGN and their twins when comparing the radial stellar angular momentum of the receding and approaching sides (see also Table \ref{tab:differences_stars_gas_angular_momentum_table}), specially in the case of the unbarred AGN. 
For the two barred galaxies we find that the stellar $\lambda_{R}$ derived from the aproaching and receding sides are intercepted at around the end of the bars. These interesting results provide us with two different parameters to consider when comparing large-scale twin galaxies differing in nuclear activity that could be related to AGN triggering.

First, the difference between the stellar angular momentum derived from the receding and approaching sides, tentatively larger in AGN and more importantly, in unbarred AGN. Such differences could be pointing to dynamical lopsidedness as an AGN triggering mechanism. Since bars naturally promote the inflow of gas towards the center, unbarred galaxies would need a more important dynamical lopsidedness to reach similar inflowing rates than their barred twins. A relation between morphological and/or kinematical lopsidedness in galaxies and nuclear activity has been already claimed (e.g. \citealt{2008ApJ...677..186R, 2009ApJ...691.1005R}). Indeed NGC2916 has already been identified as morphologically lopsided \citep{2000ApJ...538..569R} and here we confirm this kinematically. The dynamically lopsided stellar component in NGC2916 is probably related with the internal twist measured in the ionised gas disk (see Section \ref{inclina}).

Second, the stellar angular momentum $\lambda_{Re}$, tentatively larger in AGN than in their non-active twins. We suggest that the difference between the stellar $\lambda_{Re}$ measured for each AGN and its corresponding non-active twin could be the imprint of the inflow of gas that triggered the nuclear activity. In such scenario, a disk instability (e.g. a kinematically lopsided disk) would induce gas inflows from large-scales into the circumnuclear region. This inflowing gas would have transferred its angular momentum to the baryonic matter. As result, the stellar component at large-scales would increase its angular momentum, as we see in Figure \ref{fig:angular_momentum}. Recent theoretical simulations \citep{Saha_2014} link the evolution of lopsided galaxies with a smooth angular momentum transfer outwards which facilitates gas infall.

\begin{table}
\begin{center}
\begin{tabular}{ccc}
\hline
Galaxy & gaseous $\lambda_{Re}$ & stellar $\lambda_{Re}$ \\
\hline
NGC0214 (AGN-bar)  & 0.82 $\pm$ 0.01 & 0.66 $\pm$ 0.04 \\
NGC2253 (SF-bar)   & 0.75 $\pm$ 0.01 & 0.45 $\pm$ 0.04 \\
NGC2916 (AGN-nbar) & 0.85 $\pm$ 0.01 & 0.63 $\pm$ 0.04 \\
NGC0001 (SF-nbar)  & 0.79 $\pm$ 0.01 & 0.43 $\pm$ 0.02 \\
\hline
\end{tabular}
\caption[Angular momentum]{Summary of the values obtained for the specific angular momentum integrated within one effective radius.}\label{tab:angular_momentum_table}
\end{center}
\end{table}

\begin{table*}
\begin{center}
\begin{tabular}{ccccccc}
\hline
 & Galaxy & Total & 2-4\,Kpc  & 4-6\,Kpc  & 6-8\,Kpc & 8-10\,Kpc\\
\hline
\parbox[t]{2mm}{\multirow{4}{*}{\rotatebox[origin=c]{90}{Stellar}}}
 & NGC0214 (AGN-bar) & 0.06 $\pm$ 0.01 & 0.04 $\pm$ 0.01 & 0.02 $\pm$ 0.01 & 0.05 $\pm$ 0.01 & 0.12 $\pm$ 0.01 \\
 
 & NGC2253 (SF-bar)  & 0.05 $\pm$ 0.01 & 0.08 $\pm$ 0.01 & 0.02 $\pm$ 0.01 & 0.02 $\pm$ 0.01 & 0.05 $\pm$ 0.01 \\
 
 & NGC2916 (AGN-nbar) & 0.08 $\pm$ 0.01 & 0.07 $\pm$ 0.02 & 0.11 $\pm$ 0.01 & 0.12 $\pm$ 0.01 & 0.10 $\pm$ 0.01 \\
 
  & NGC0001 (SF-nbar) & 0.04 $\pm$ 0.01 & 0.02 $\pm$ 0.01 & 0.02 $\pm$ 0.01 & 0.05 $\pm$ 0.01 & 0.06 $\pm$ 0.01 \\
\hline
\parbox[t]{2mm}{\multirow{4}{*}{\rotatebox[origin=c]{90}{Gas}}} 
 & NGC0214 (AGN-bar) & 0.05 $\pm$ 0.01 & 0.05 $\pm$ 0.01 & 0.02 $\pm$ 0.01 & 0.01 $\pm$ 0.01 & 0.05 $\pm$ 0.02 \\
 
 & NGC2253 (SF-bar)  & 0.09 $\pm$ 0.02 & 0.19 $\pm$ 0.02 & 0.05 $\pm$ 0.01 & 0.01 $\pm$ 0.01 & 0.01 $\pm$ 0.01 \\
 
 & NGC2916 (AGN-nbar) & 0.10 $\pm$ 0.03 & 0.23 $\pm$ 0.03 & 0.12 $\pm$ 0.05 & 0.01 $\pm$ 0.01 & 0.01 $\pm$ 0.01 \\
 
 & NGC0001 (SF-nbar) & 0.05 $\pm$ 0.01 & 0.03 $\pm$ 0.01  & 0.01 $\pm$ 0.01 & 0.03 $\pm$ 0.01 & 0.02 $\pm$ 0.01 \\
\hline

\end{tabular}
\caption[]{Differences between the stellar and gaseous $\lambda_{R}$  radial profiles for the approaching and receding sides.}\label{tab:differences_stars_gas_angular_momentum_table}
\end{center}
\end{table*}

\subsection{Stellar populations}\label{sec:effects_stellar_population}

\begin{figure*}
    \centering   
    \includegraphics[width=\textwidth]{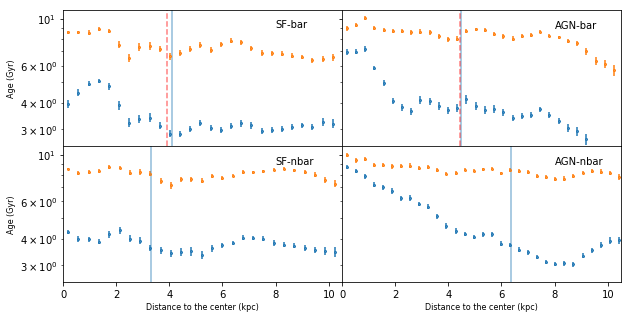}    
    \caption{Mean luminosity (blue) and mass (orange) weighted age gradients as a function of radius for the four galaxies. Blue solid lines indicate the effective radius and red dashed lines indicate the radius of the bars.}
    \label{fig:age_Tomas_arcs}
\end{figure*}

\begin{figure*}
    \centering
    \includegraphics[width=\textwidth]{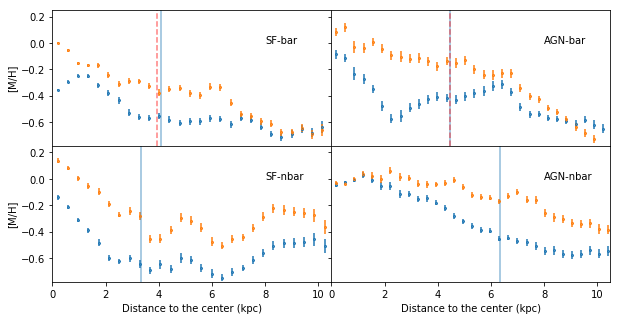}   
    \caption{Mean luminosity (blue) and mass (orange) metallicity gradients as a function of radius for the sample of our galaxies. Blue solid lines indicate the effective radius and red dashed lines indicate the radius of the bars.}
    \label{fig:metallicity_Tomas_arcs}
\end{figure*}

\begin{figure*}
    \centering
    \includegraphics[width=\textwidth]
{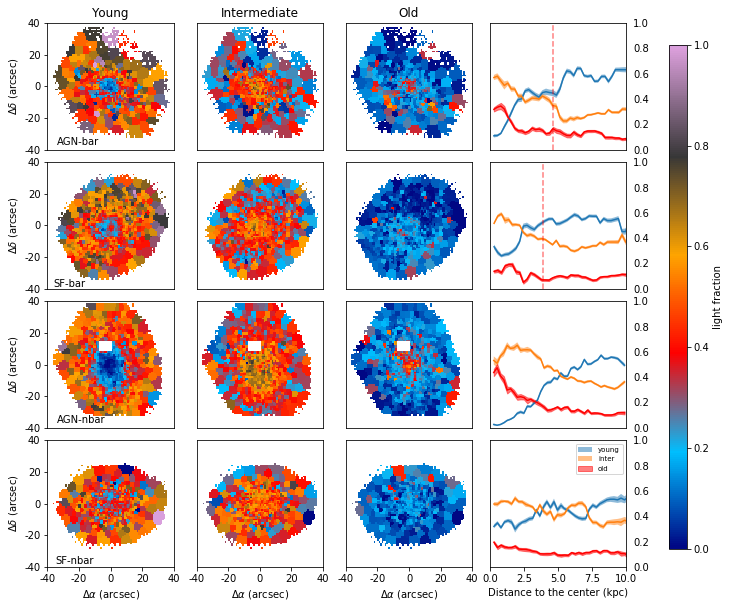}
    \caption{Maps showing the relative contribution of young (blue), intermediate (orange) and old (red)  stellar populations in each spatial element throughout the galaxies and radial profiles of each component. Red dashed lines indicate the radius of the bars.}
    \label{fig:maps_profiles_SFH_Tomas}
\end{figure*}

\begin{table*}
\begin{center}
\begin{tabular}{ccccccc}
\hline
Galaxy &  & Central (<1\,kpc)  &   &  & Inner (1<r<5)\,kpc &  \\
 & Young & Intermediate  & Old  & Young & Intermediate & Old \\
\hline
NGC0214 (AGN-bar)  & 11 $\pm$ 1 & 56 $\pm$ 1 &  33 $\pm$ 1 & 43 $\pm$ 2 & 41 $\pm$ 1 & 16 $\pm$ 1 \\

NGC2253 (SF-bar)  &  29 $\pm$ 2 & 57 $\pm$ 2 & 14 $\pm$ 1 &  49 $\pm$ 3 & 42 $\pm$ 2 & 10 $\pm$ 1 \\

NGC2916 (AGN-nbar) & 3  $\pm$ 1 & 55 $\pm$ 2 & 42 $\pm$ 2 & 18  $\pm$ 3 & 60 $\pm$ 2 & 21 $\pm$ 1 \\

NGC0001 (SF-nbar)  & 32 $\pm$ 1 & 50 $\pm$ 1 & 17 $\pm$ 1 &  38 $\pm$ 2 & 49 $\pm$ 1 & 13 $\pm$ 1 \\
\hline

\end{tabular}
\caption[]{Light fractions in percentages in the central (1\,kpc) and inner (1 < r < 5\,kpc) parts of young (< 1.5\,Gyr), intermediate (1.5\,Gyr < intermediate < 10\,Gyr) and old (>10\,Gyr) populations.}\label{tab:table_maps_profiles_light_weighted}
\end{center}
\end{table*}

Figures \ref{fig:age_Tomas_arcs} and \ref{fig:metallicity_Tomas_arcs} show the age and metallicity gradients of the modelled stellar populations as a function of radius for the four galaxies.
The SF-nbar galaxy presents a flat age profile, while its barred twin (i.e. the SF-bar) presents a younger and less metallic stellar population at the center and at the edges of the bar than along it. This result suggests that the bar could have driven material from the disk, of lower metallicity, towards the centre of the galaxy, feeding a central star formation episode. Previous observational (e.g. \citealt{2017MNRAS.470L.122P}) and theoretical (e.g. \citealt{2007A&A...465L...1W}) works support the presence of age gradients along stellar bars in galaxies, with younger stellar populations at the center and bar edges. Furthermore, while the SF-nbar shows a flat age profile, its active twin shows younger stellar populations outwards, reaching similar ages beyond 6\,kpc. On the contrary, the barred twins differing in nuclear activity follow similar age profiles. Interestingly, both AGN show an older stellar population at the nuclear region than their non-active twins (see Fig. \ref{fig:metallicity_Tomas_arcs}).

We have also computed the mean age (luminosity-weighted) within the effective radius. We found that both AGN present older average stellar populations than their SF twins, in agreement with previous studies \citep{2003MNRAS.346.1055K, 2017MNRAS.472.4382R}, at least when low-luminosity AGN are considered. 

Moreover, we have also analysed the spatial and radial distribution of three different stellar sub-populations (see Sec. \ref{Stellar population method} and Fig. \ref{fig:maps_profiles_SFH_Tomas}). We find that the central region (< 1\,kpc) of the two AGN are dominated by the light of an intermediate stellar population with almost no young stellar population (age < 1.5\,Gyr; 11\% and 3\% for the AGN-bar and the AGN-nbar, respectively). On the contrary, young stellar populations contribute up to 30\% at the central region of their non-active twins. Additionally, we find a larger contribution from old stellar populations (age > 10\,Gyr) in the central 5\,kpc of the two AGN than in their non-active twins (see Table \ref{tab:table_maps_profiles_light_weighted}).

The connection  between nuclear activity and star formation is still an open question but they both require a cold gas supply. The effectiveness of the fuelling, the regulation of both processes and their relation have important implications for the growth of galaxies and black holes over cosmic time. Following our results we propose that part of the gas that should have contributed to maintain star formation could have inflowed to the central region of the active galaxies, driven by a disk instability, a bar or a combination of the two.

\section{Conclusions}\label{sec:Conclusions}

Taking advantage of the CALIFA Survey, we present a comparison of two barred and unbarred pairs of almost identical twin galaxies differing in nuclear activity. We mainly focus on identifying the large-scale parameters that could be related to AGN triggering. The main results and conclusions of this pilot study are the following: 

\begin{itemize}
\item The stellar and ionised gas velocity fields of the four galaxies can be reproduced with a simple rotating disk at kpc-scales. However, we detect an internal twist in the ionised gas disk of the unbarred AGN that could be playing the same role as the bar of the barred AGN in driving an inflow of gas to the central region.

\item The studied active galaxies show higher values of $\lambda_{Re}$ and smaller differences between the stellar and ionised gas $\lambda_{Re}$ than their non-active twins, suggesting that their stellar disks could have gained part of their angular momentum from the inflowing gas that triggered the AGN.

\item The radial profiles of the stellar angular momentum derived from the receding and approaching sides of the velocity fields of the two AGN show larger differences than those of their non-active twins. This suggests a dynamical lopsidedness in the AGN that appears larger in the unbarred AGN. 

\item The central regions of the active galaxies show a smaller contribution from young stellar populations than their non-active twins. The dynamical lopsidedness found in the active galaxies could have produced the infall of gas that triggered nuclear activity and prevent nuclear star formation. 

\item Regardless of having an AGN, the analysed barred galaxies show smaller velocity dispersions at the centre than their unbarred twins. This finding could be indicating that barred galaxies could have followed slightly different evolutionary paths than their unbarred twins in spite of having a similar large-scale appearance.

\item The barred galaxies present lower central metallicities than their unbarred twins regardless of having nuclear activity, suggesting that the bar has driven material from the disk inwards.

\end{itemize}

Through this pilot study we have identified four kinematic parameters at kpc-scales that differ when we compare large-scale identical twin galaxies differing in AGN or bar presence:
\begin{itemize}

\item The difference in inclination derived from the approaching and receding sides of the velocity field, suggestive of an internal twist in unbarred AGN compared to its non-active  twin. 

\item The stellar angular momentum integrated within the effective radius, found to be larger in AGN than in their non-active twins.

\item The difference in the stellar angular momentum radial profiles derived from the approaching and receding sides of the velocity field, found larger in AGN than in their non-active twins.

\item The relative contribution from young stellar populations in the central regions, found to be lower in AGN than in their non-active twins.
\end{itemize}

These four parameters could be imprints of the gas inflow from large-scale to the nuclear region that triggered AGN. The next step is to explore statistically these parameters and confirm or discard the proposed scenario.

\section*{Acknowledgements}


This study uses data provided by the Calar Alto Legacy Integral Field Area (CALIFA) survey (\url{http://califa.caha.es/}). Based on observations collected at the Centro Astron\'omico Hispano Alem\'an (CAHA) at Calar Alto, operated jointly by the Max-Planck-Institut f{\"u}r Astronomie and the Instituto de Astrof\'isica de Andaluc\'ia (CSIC). IMC acknowledges the support of the Instituto de Astrof\'isica de Canarias via an Astrophysicist Resident fellowship.  B.G-L acknowledges support from the Spanish Ministerio de Economia y Competitividad (MINECO) by the grant AYA2015-68217-P. CRA acknowledges the Ram\'on y Cajal Program of the Spanish Ministry of Economy and Competitiveness through project RYC-2014-15779 and the Spanish Plan Nacional de Astronom\'ia y Astrofis\'ica under grant AYA2016-76682-C3-2-P. TRL acknowledges support via grants AYA2014-56795-P and AYA2016-77237-C3-1-P from the Spanish Government. SFS thanks the CONACyT programs CB-285080 and DGAPA IA101217 grants for their support. IMP and JM acknowledges support from the Spanish Ministerio de Economia y Competitividad (MINECO) by the grant AYA2016-76682-C3-1-P. This paper makes use of python (\url{http://www.python.org}); Matplotlib \citep{Hunter_2007}, a suite of open-source python modules that provide a framework for creating scientific plots; and Astropy,\footnote{\url{http://www.astropy.org}} a community-developed core Python package for Astronomy \citep{Astropy_2013, Astropy_2018}. Finally, we thank the anonymous referee for useful suggestions and comments that have improved this paper.




\bibliographystyle{mn2e}
\bibliography{biblio} 




\appendix

\section{Notes on individual objects}\label{sec:kinematic_map}

The first part of this appendix is devoted to describe the characteristics of the individual objects. In the second part we present the different maps obtained for each galaxy.

\subsection{NGC0214 (AGN-bar)}

\begin{figure*}
    \centering
    \includegraphics[width=.76\textwidth]{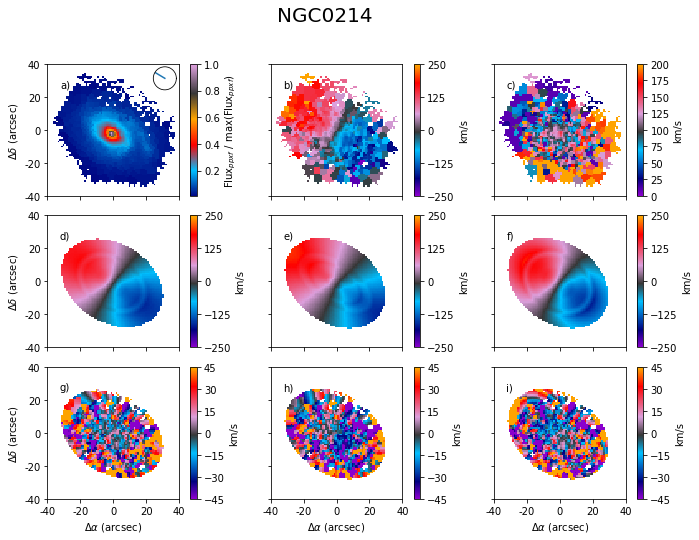}
    \caption{NGC0214 (AGN-bar): Stellar flux (a), velocity (b) and velocity dipersion (c) maps. The circle at the top-right corner of the flux map indicates the bar orientation. Disk models fitted to the total (d), approaching (e) and receding sides (f) of the velocity field. Residuals after subtracting the models fitted to the total (g), approaching (h) and receding (i) sides of the velocity field from the total velocity field.}
    \label{fig:NGC0214_stars}
\end{figure*}

\begin{figure*}
    \centering
    \includegraphics[width=.75\textwidth]{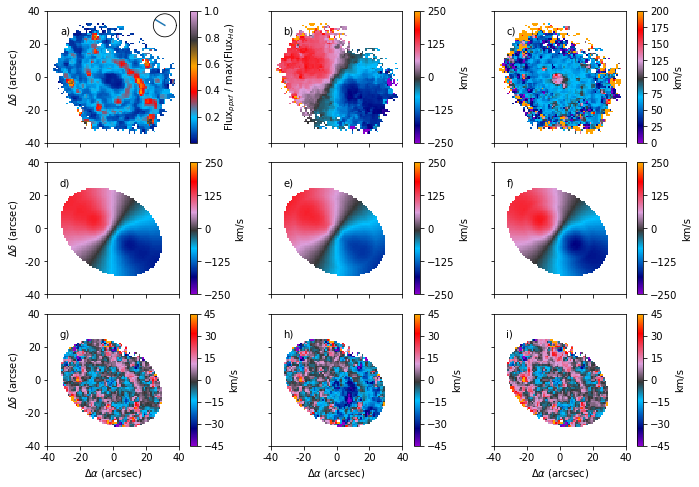}
    \caption{Same as in Fig. \ref{fig:NGC0214_stars} but for the ionised gas component of NGC0214 (AGN-bar)}
    \label{fig:NGC0214_ha}
\end{figure*}

The stellar continuum map  (top left panel of Fig. \ref{fig:NGC0214_stars}) reflects the elongation of the bar in the inner region. The H$\alpha$ map (top left panel of Fig. \ref{fig:NGC0214_ha}) shows a patchy distribution with weak emission in the central $\sim$12\,arcsec (3\,kpc) of the galaxy and stronger emission in the spiral arms. 

The global photometric and stellar/gaseous kinematic PA are in agreement. While the kinematic PA obtained from the receding and approaching sides are also in agreement, the inclinations differ around 3 degrees (Table \ref{tab:big_table}). This difference in inclination is reduced when we only use the ionised gas velocities at large galactocentric distances (Table \ref{tab:big_table_mask}).

\begin{figure*}
    \centering
    \includegraphics[width=.76\textwidth]{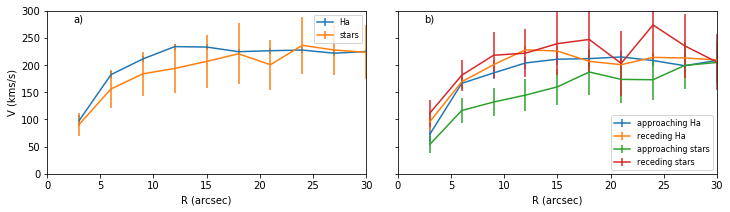}
    \caption{a) Ionised gas (blue line) and stellar (orange line) rotation curves derived from the simple disk model. b) Rotation curves derived from the fits to the approaching and receding sides of the velocity field of NGC0214 (AGN-bar).}
    \label{fig:NGC0214_velocity_curves}
\end{figure*}

The stellar and ionised gas rotation curves (Fig. \ref{fig:NGC0214_velocity_curves}) show similar behaviours within the uncertainties. The stellar rotation velocities derived from the approaching side are slightly smaller than those measured from the receding side.

When we subtracted the simple disk model from the observed velocity field we find positive residuals along the west spiral arm, the brightest in H$\alpha$ (last column of Fig. \ref{fig:NGC0214_ha}). 
For the residual map obtained when subtracting the disk model from the approaching side of the ionised gas velocity field (bottom middle panel of Fig. \ref{fig:NGC0214_ha}), regions of strong negative residuals appear around the nucleus and at 10 arcsec south-west. 

The light-weighted age radial profile shows an older population towards the center (see Figure \ref{fig:age_Tomas_arcs}). However, the central kpc shows an almost flat behaviour. The light-weighted metallicity (Fig. \ref{fig:metallicity_Tomas_arcs}) also presents a decreasing gradient from the center. The central region of NGC0214 is dominated by the light of intermediate age (56$\%$) and old (33$\%$) stellar populations (see Table \ref{tab:table_maps_profiles_light_weighted} and Fig. \ref{fig:maps_profiles_SFH_Tomas}), while the contribution of the young component increases outwards (11\,$\%$ in the central kpc and 43\,$\%$ between 1 and 5\,kpc).

\subsection{NGC2253 (SF-bar)}

The stellar continuum map clearly shows the bar elongation (top left panel of Fig. \ref{fig:NGC2253_stars}). The H$\alpha$ map reveals a faint circumnuclear ring at 1-2\,kpc from the nucleus (top left panel of Fig. \ref{fig:NGC2253_ha}). 

NGC2253 shows a morpho-kinematic PA misalignment of around 10 degrees. The stellar and ionised gas kinematic parameters are in good agreement (see Table \ref{tab:big_table} and \ref{tab:big_table_mask}).
The stellar rotation curves (total, approaching and receding) fall below the ionised gas ones, suggesting that a pressure supported contribution to the stellar kinematics is not negligible in NGC2253 (Fig. \ref{fig:NGC2253_velocity_curves}).

The ionised gas residual maps (last column of Fig. \ref{fig:NGC2253_ha}) present positive residual velocities in almost all the bright H$\alpha$ knots. 

The radial profiles of the light-weighted age/metallicity  (Figs \ref{fig:age_Tomas_arcs} and \ref{fig:metallicity_Tomas_arcs}) reveal a younger and less metallic stellar population in the central kpc and at the edges of the bar than  along it, with a flat gradient along the disk. Maps of stellar sub-populations  (Fig. 
\ref{fig:maps_profiles_SFH_Tomas}) reproduce such behaviour, showing a circumnuclear young population-free ring on the faint H$\alpha$ ring structure.

\subsection{NGC2916 (AGN-nbar)}

The stellar continuum map (top left panel of Fig. \ref{fig:NGC2916_stars}) peaks at the nucleus and the H$\alpha$ map (top left panel of Fig. \ref{fig:NGC2916_ha}) shows weak emission in the central 3-4\,kpc and stronger patchy emission tracing the spiral arms.

Both the stellar and ionised gas velocity fields are dominated by rotation at large scales (to middle panels of Figs. \ref{fig:NGC2916_stars} and \ref{fig:NGC2916_ha}). However, the H$\alpha$ map shows a disturbed circumnuclear region with high velocity dispersions likely due to the poor S/N. 

Global photometric and kinematic PA are in agreement. A difference  of about 6 degrees is derived when comparing the inclinations obtained from fitting the disk model to the receding or approaching sides of the ionised gas velocity field (see Table \ref{tab:big_table}). This difference in inclination remains when the ionised gas velocities at large galactocentric distances are only considered for the model fit.

The rotation curves follow a similar behaviour for the stellar and ionised gas components within the errors (see Fig. \ref{fig:NGC2916_velocity_curves}). 
The faint H$\alpha$ emission in the circumnuclear region provides the strongest velocity residuals (last column of Fig. \ref{fig:NGC2916_ha}).

The light-weighted age radial profile (see Fig. \ref{fig:age_Tomas_arcs}) shows an older population towards the center. The light-weighted metallicity (see Fig. \ref{fig:metallicity_Tomas_arcs}) also decreases outwards. The profiles in Fig.  \ref{fig:maps_profiles_SFH_Tomas} and Table \ref{tab:table_maps_profiles_light_weighted} indicate that this galaxy is dominated in the center by intermediate (55\,$\%$) and old (42\,$\%$) stellar populations with a small contribution of a young population (3\,$\%$). The young population increases with the galactocentric distance while the intermediate and old populations decrease.

\subsection{NGC0001 (SF-nbar)}

The stellar continuum map (top left panel of Fig. \ref{fig:NGC0001_stars}) peaks at the center of the galaxy, while the H$\alpha$ map shows emission in the spiral arms (top left panel of Fig. \ref{fig:NGC0001_ha}).
The kinematic parameters derived from the approaching and receding sides are in good agreement within the estimated uncertainties (see Table \ref{tab:big_table} and \ref{tab:big_table_mask}).

The stellar rotation curves (from total, receding and approaching sides) fall below the ionised gas rotation curves  and the stellar velocity dispersion is higher than the gaseous velocity dispersion. This behaviour suggests that the stars could be partly supported by pressure \citep{Vega_Beltran_2001,Beckman_2004}. 
Beyond 15\,arcsec, only the rotation curve of the ionised gas can be obtained showing a flat behaviour (see Fig. \ref{fig:NGC0001_velocity_curves}).

The velocity residual maps show patchy structures for both the stellar and ionised gas kinematics (last column of Fig. \ref{fig:NGC0001_ha}).

The radial behavior of the light-weighted age (see Fig. \ref{fig:age_Tomas_arcs}) in NGC0001 is quite flat,  while the light weighted metallicity (see Fig. \ref{fig:metallicity_Tomas_arcs}) shows high values in the nucleus of the galaxy, then decreases until the R$_{e}$ and then remains flat. 

\begin{figure*}
    \centering
    \includegraphics[width=0.7\textwidth]{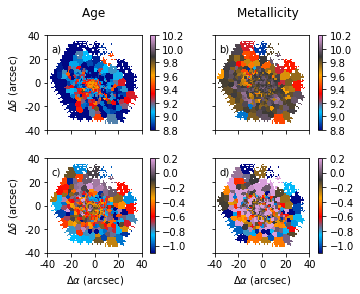}
    \caption{a) Luminosity-weighted log(age(yr)) map, b) mass-weighted log(age(yr)) map, c) luminosity-weighted [M/H] map and d) mass-weighted [M/H] map for NGC0214 (AGN-bar).}   
    \label{fig:NGC0214_maps_Tomas}
\end{figure*}

\begin{figure*}
    \centering
    \includegraphics[width=.76\textwidth]{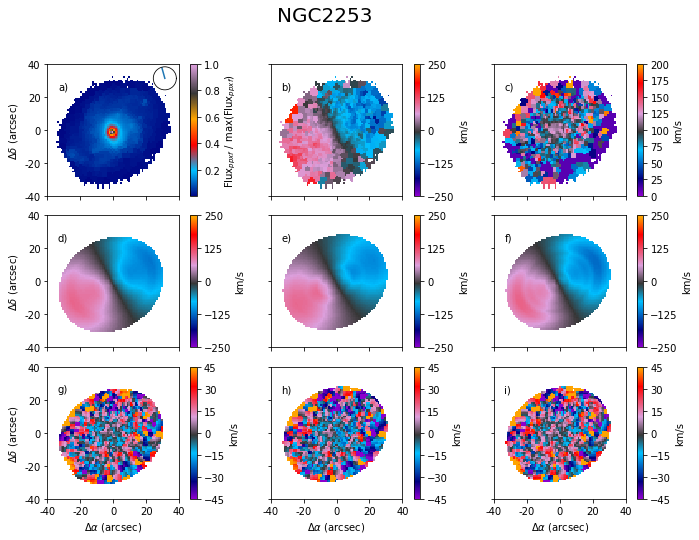}
    \caption{Same as in Fig. \ref{fig:NGC0214_stars} but for NGC2253 (SF-bar).}
    \label{fig:NGC2253_stars}
\end{figure*}

\begin{figure*}
    \centering
    \includegraphics[width=.76\textwidth]{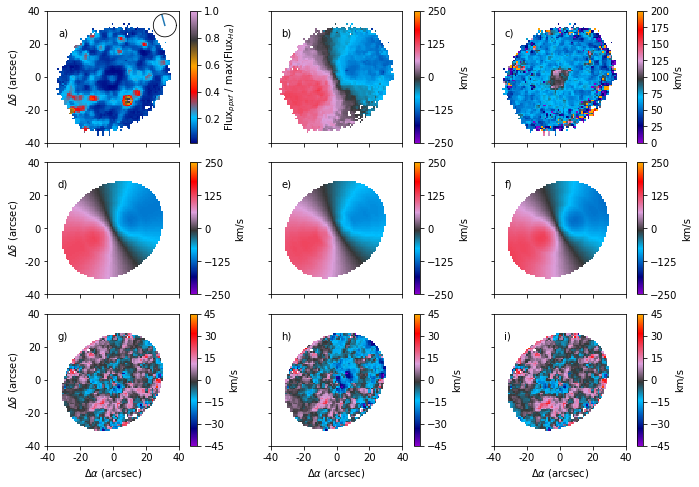}
    \caption{Same as in Fig. \ref{fig:NGC0214_ha} but for NGC2253 (SF-bar).}
    \label{fig:NGC2253_ha}
\end{figure*}

\begin{figure*}
    \centering
    \includegraphics[width=.7\textwidth]{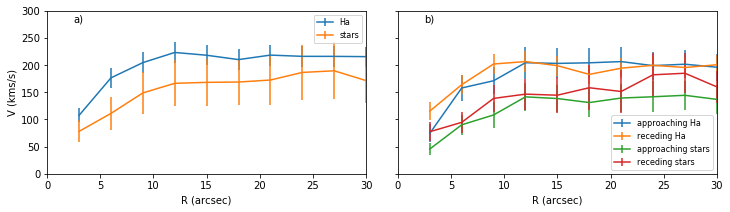}
    \caption{Same as in Fig. \ref{fig:NGC0214_velocity_curves} but for NGC2253 (SF-bar).}
    \label{fig:NGC2253_velocity_curves}
\end{figure*}

\begin{figure*}
    \centering
    \includegraphics[width=.7\textwidth]{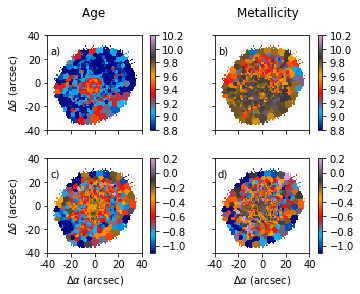}
    \caption{Same as in Fig. \ref{fig:NGC0214_maps_Tomas} but for NGC2253 (SF-bar).}
    \label{fig:NGC2253_maps_Tomas}
\end{figure*}

\begin{figure*}
    \centering
    \includegraphics[width=.76\textwidth]{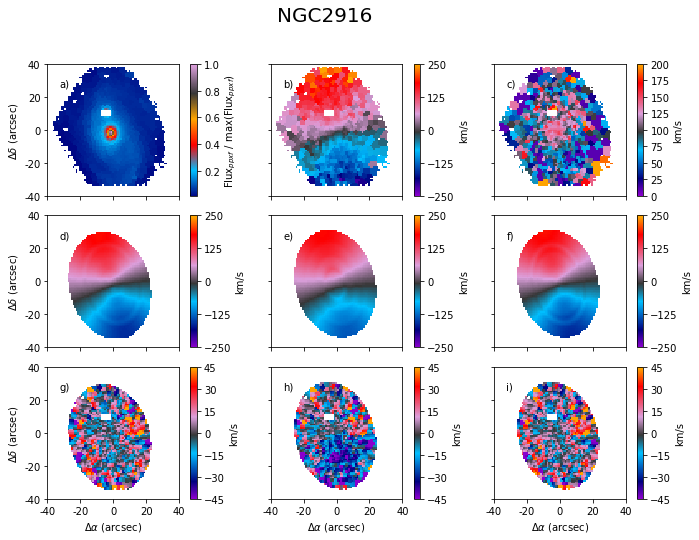}
    \caption{Same as in Fig. \ref{fig:NGC0214_stars} but for NGC2916 (AGN-nbar).}
    \label{fig:NGC2916_stars}
\end{figure*}

\begin{figure*}
    \centering
    \includegraphics[width=.76\textwidth]{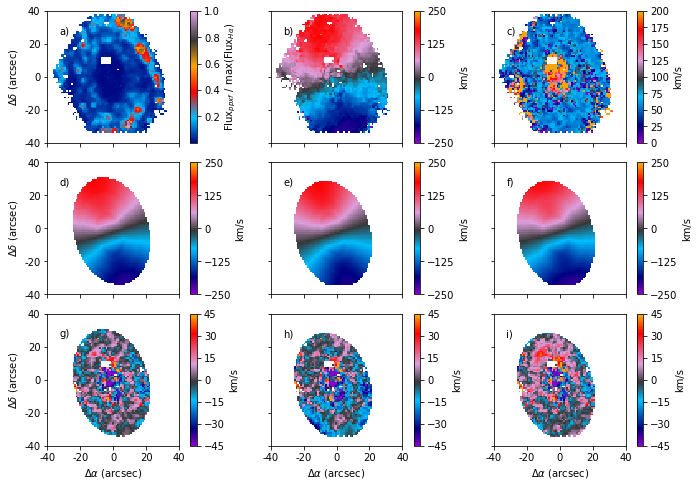}
    \caption{Same as in Fig. \ref{fig:NGC0214_ha} but for NGC2916 (AGN-nbar).}
    \label{fig:NGC2916_ha}
\end{figure*}

\begin{figure*}
    \centering
    \includegraphics[width=.7\textwidth]{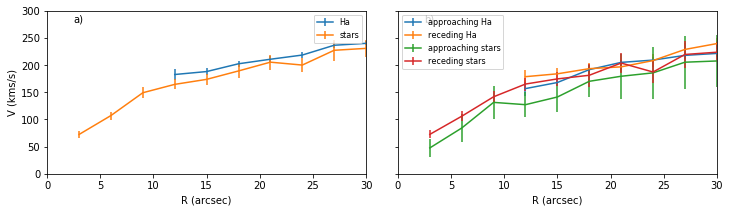}
    \caption{Same as in Fig. \ref{fig:NGC0214_velocity_curves} but for NGC2916 (AGN-nbar).}
    \label{fig:NGC2916_velocity_curves}
\end{figure*}

\begin{figure*}
    \centering
    \includegraphics[width=.7\textwidth]{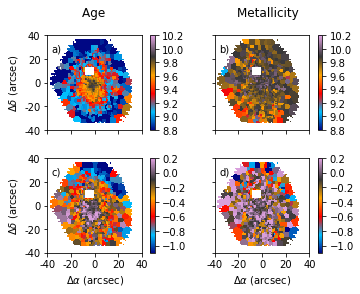}
    \caption{Same as in Fig. \ref{fig:NGC0214_maps_Tomas} but for NGC2916 (AGN-nbar).}
    \label{fig:NGC2916_maps_Tomas}
\end{figure*}

\begin{figure*}
    \centering
    \includegraphics[width=.76\textwidth]{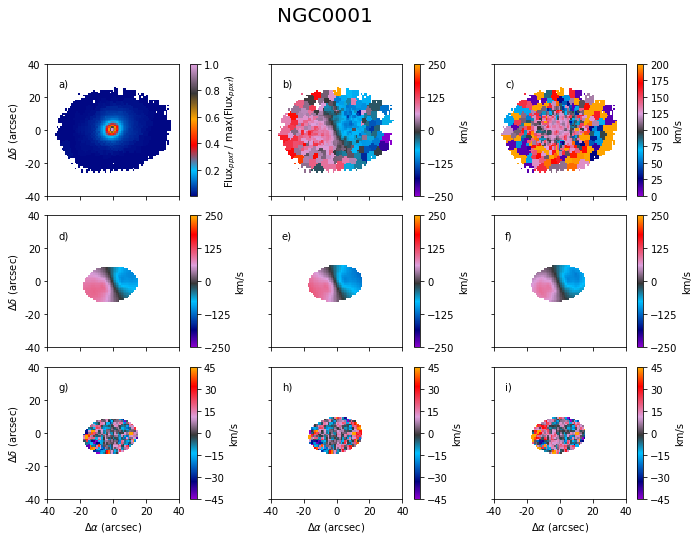}
    \caption{Same as in Fig. \ref{fig:NGC0214_stars} but for NGC0001 (SF-nbar).}
    \label{fig:NGC0001_stars}
\end{figure*}

\begin{figure*}
    \centering
    \includegraphics[width=.76\textwidth]{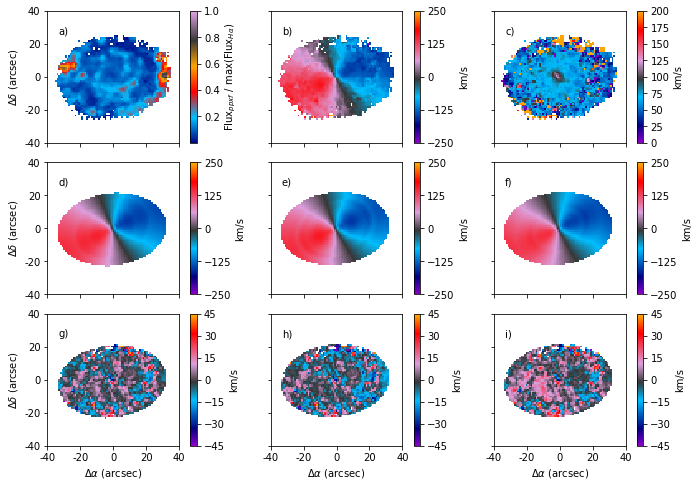}
    \caption{Same as in Fig. \ref{fig:NGC0214_ha} but for NGC0001 (SF-nbar).}
    \label{fig:NGC0001_ha}
\end{figure*}

\begin{figure*}
    \centering
    \includegraphics[width=.7\textwidth]{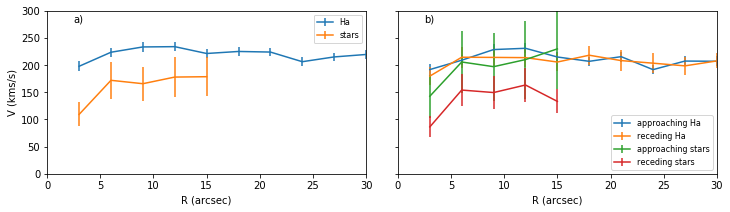}
    \caption{Same as in Fig. \ref{fig:NGC0214_velocity_curves} but for NGC0001 (SF-nbar).}
    \label{fig:NGC0001_velocity_curves}
\end{figure*}

\begin{figure*}
    \centering
    \includegraphics[width=.7\textwidth]{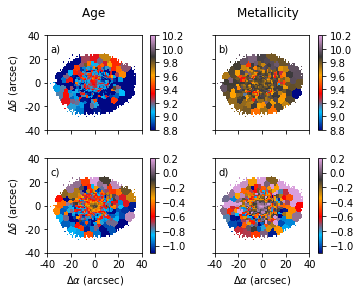}
    \caption{Same as in Fig. \ref{fig:NGC0214_maps_Tomas} but for NGC0001 (SF-nbar).}
    \label{fig:NGC0001_maps_Tomas}
\end{figure*}


\bsp	
\label{lastpage}
\end{document}